\documentclass{elsart}

\usepackage{graphicx}
\usepackage{dcolumn}
\usepackage{bm}

\begin{document}
\begin{frontmatter}
\begin{flushright}
{\bf ICRR-Report-484-2002-2}
\end{flushright}

\title{
Detecting    
Very High Energy Neutrinos \\
by the Telescope Array
}

\author{Makoto Sasaki\thanksref{a1}}
 \thanks[a1]{sasakim@icrr.u-tokyo.ac.jp}
\author{Yoichi Asaoka\thanksref{a2}}
 \thanks[a2]{asaoka@icrr.u-tokyo.ac.jp}
\author{Masashi Jobashi\thanksref{a3}}
 \thanks[a3]{jobashi@icrr.u-tokyo.ac.jp}
\address{%
Institute for Cosmic Ray Research, University of Tokyo, 5-1-5 Kashiwanoha, Kashiwa 277-8582, Japan \\
}%

\date{June 10, 2002}

\begin{abstract}
We present the cosmic neutrino detecting
potential of the Telescope Array for
quasi-horizontally downward and upward air showers initiated 
by very high energy neutrinos penetrating the air and the Earth respectively.
We adopted model predictions for extraterrestrial neutrino fluxes 
from active galactic nuclei, gamma-ray bursts, Greisen photo production,  
the collapse of topological defects, and Z-bursts.
The Telescope Array, 
using a large array of bright and wide field-of-view fluorescence telescopes,
can explore these very high energy cosmic neutrino sources.
\end{abstract}

\end{frontmatter}

\section{Introduction}
The detection of very high energy cosmic neutrinos is important from 
at least three points of view.
First, 
they can carry astrophysical information about point sources 
which are optically thick at almost all wavelengths
and relics of an early inflationary phase in the history of the universe.
Second, the connection between the emission of cosmic nuclei, gamma rays, and 
neutrinos from astrophysical accelerators is of considerable interest
with regards 
the problem of the origin of extragalactic 
extremely high energy cosmic rays.
Third, 
due to vacuum oscillations, 
a flux of high energy cosmic neutrinos 
is  
expected to be 
almost equally distributed among the three flavors due to vacuum oscillations, 
provided that the neutrinos are produced by 
distant objects 
such as gamma ray bursts (GRBs)
and active galactic nuclei (AGN), 
and hence the
study of oscillation effects on high energy neutrino 
fluxes 
can 
be used to study
neutrino mixing and 
distinguish between different mass schemes~\cite{bib:Athar00}.


In this paper we present the potential of the 
Telescope Array (TA), which will be able to detect
cosmic 
neutrinos with energies above 10$^{16}$~eV, 
so far unobserved. 
They
have great potential as probes of astrophysics and particle physics phenomena.
They have escaped from regions of dense matter and point back to their sources 
and
they provide a unique window into the most violent events in the universe.
Hadrons are accelerated to extremely high energies, regardless of their 
source, accelerating mechanism, and
composition (i.e. $pp$ and/or $p\gamma\rightarrow\pi^{\pm}$), 
and a subsequent decay of accompanying pions is expected to 
result in neutrino fluxes through the decay process 
$\pi\rightarrow\mu + \nu_{\mu}\rightarrow e + \nu_e \nu_{\mu} + \nu_{\mu}$, 
with a resultant ratio $\nu_{\mu} / \nu_{e}$=2. 
Observations 
~\cite{bib:agasa-spect,bib:flyseye-highest,bib:yakutsk-highest}
of cosmic rays with energies beyond the Greisen-Zatsepin-Kuzmin (GZK) cutoff  
~\cite{bib:Greisen66,bib:ZatsepinKuzmin66}
raise interest in 
extremely high energy neutrinos.
The GZK mechanism can lead to neutrino fluxes well above 
those from proposed exotic sources for the super-GZK events
around the energy of 10$^{18}$~eV.

\begin{figure}[b!]
\begin{center}
\includegraphics[width=0.95\linewidth]{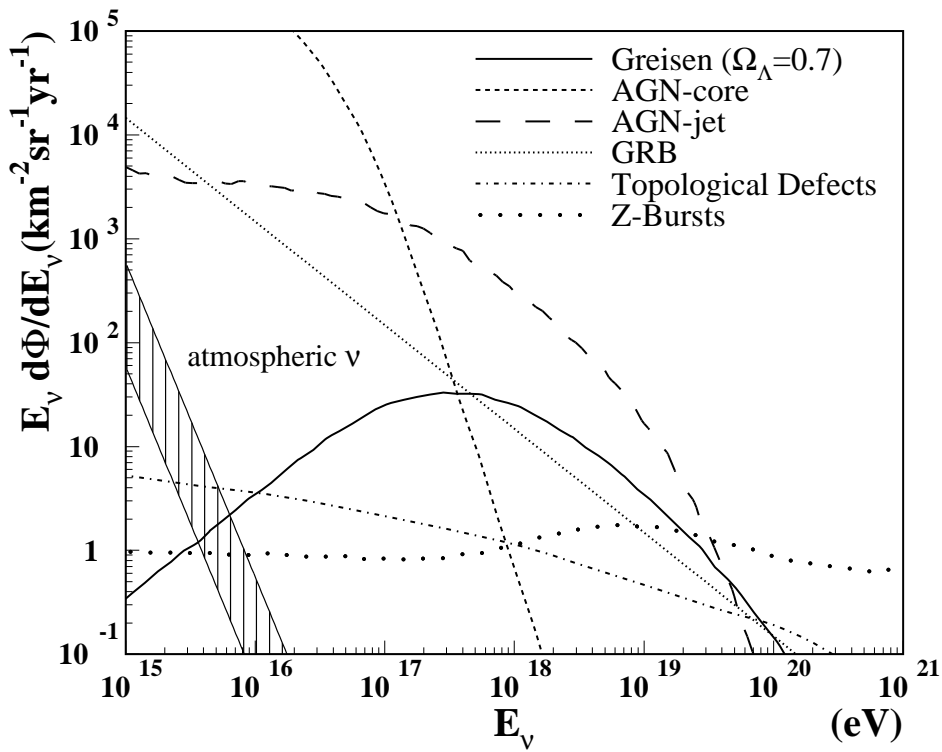}
\caption{
Differential fluxes of muon neutrinos ($\nu_{\mu}+\bar{\nu}_{\mu}$)
from Greisen photo production~\cite{bib:Engel01},
active galactic nuclei~\cite{bib:Mannheim95,bib:SS95},
gamma ray bursts~\cite{bib:Waxman95},
topological defects~\cite{bib:Sigl97},
and Z-bursts~\cite{bib:Yoshida98}.
An AGN-core flux model~\cite{bib:SZ-P92} has been rejected
by AMANDA-B10~\cite{bib:AMANDA02}. 
%
%
The band with vertical hatching shows the range of atmospheric
neutrino background~\cite{bib:Volkova01} as the zenith angle changes
from 90$^\circ$ (highest) to 0$^\circ$ (lowest).
%
\label{fig:neutrino-flux}}
\end{center}
\end{figure}

A curious coincidence between the energy flow of the highest energy cosmic rays 
and gamma ray bursts (GRB) suggests a possible common source~\cite{bib:Waxman95}.
Another speculation is that very high energy cosmic rays may be 
the result of the annihilation of topological defects left over 
from the early universe~\cite{bib:Hill87,bib:Bhattacharjee92}.
The energy scales of such events are of the order of 10$^{24}$~eV (GUT scale).
In addition to these ideas is the proposal that the highest energy cosmic rays 
are evidence for novel particle physics or astrophysics~\cite{weiler99}.
These super-GZK events provide motivation for examining and confirming 
predictions of modes of photo production mechanisms.
The neutrino fluxes of some representative sources 
and the range of atmospheric neutrino background as the zenith angle changes from 
$0^{\circ}$ and $90^{\circ}$
are given in Fig.\ref{fig:neutrino-flux}.
Note the atmospheric neutrino background~\cite{bib:Volkova01}
appears to have so large spectral index
that it cannot largely contribute to the neutrino detection
in the higher energy region above 10$^{16}$~eV in any source models. 
In each of these source cases, 
neutrino induced air-showers penetrate the atmosphere to a great depth.
The imaging technique used in the TA detector has an advantage
in this respect as it can observe the slant depth of a deeply penetrating air-shower.


The recent discovery of near-maximal 
$\nu_{e}$-$\nu_{\mu}$ and 
$\nu_{\mu}$-$\nu_{\tau}$ mixing~\cite{bib:nu-mix} has 
a significant impact on the detection strategy.
The proposed astrophysical sources produce predominantly 
$\nu_{\mu}$ and $\nu_{e}$
with very small admixtures of $\nu_{\tau}$.
However, these sources are so far away that, even at the high energies of interest here,
cosmic neutrino fluxes with a $\nu_{e} : \nu_{\mu} : \nu_{\tau}$ ratio at the source 
of $1 : 2 : 0$,
inevitably oscillate in the three neutrino framework to a ratio of
$1 : 1 : 1$, irrespective of the mixing angle.

By looking for upward moving showers from Earth-skimming tau neutrinos, one can 
test for the existence of neutrino oscillations~\cite{bib:Feng01,farg99}.
Once created in the Earth, 
a charged lepton loses energy through Bremsstrahlung, pair production,
and photonuclear interactions. 
At the Earth's surface, with a density of about 2.7~g/cm$^{3}$,
tau leptons and muons are expected to travel 11~km and 1.5~km, respectively, before 
losing a decade in energy; electrons however  do not travel any significant distance.
Detectable neutrinos therefore skim the Earth at angles approximately 1$^{\circ}$
above the horizontal and these events are expected to be predominantly
showers induced by tau neutrinos. 

Another important signature of very high energy tau neutrinos is the 
``double bang'' which they produce in the atmosphere.
The initial shower is produced by the original interaction which
creates a tau lepton and a hadronic shower.
This is followed by the decay of the tau lepton, producing the second shower bang.
The two bangs are separated by a distance of 
$\sim$4.9~km~(E$_{\tau}$/10$^{17}$~eV) where E$_{\tau}$ is the energy of
the tau lepton converted from the tau neutrino.
This decay length is measurable with the imaging telescopes 
in the TA detector.

\section{The TA Detector} 
\begin{figure*}[t]
\begin{center}
\begin{minipage}{0.35\linewidth}
\includegraphics[width=\linewidth,bb=0 150 567 650]{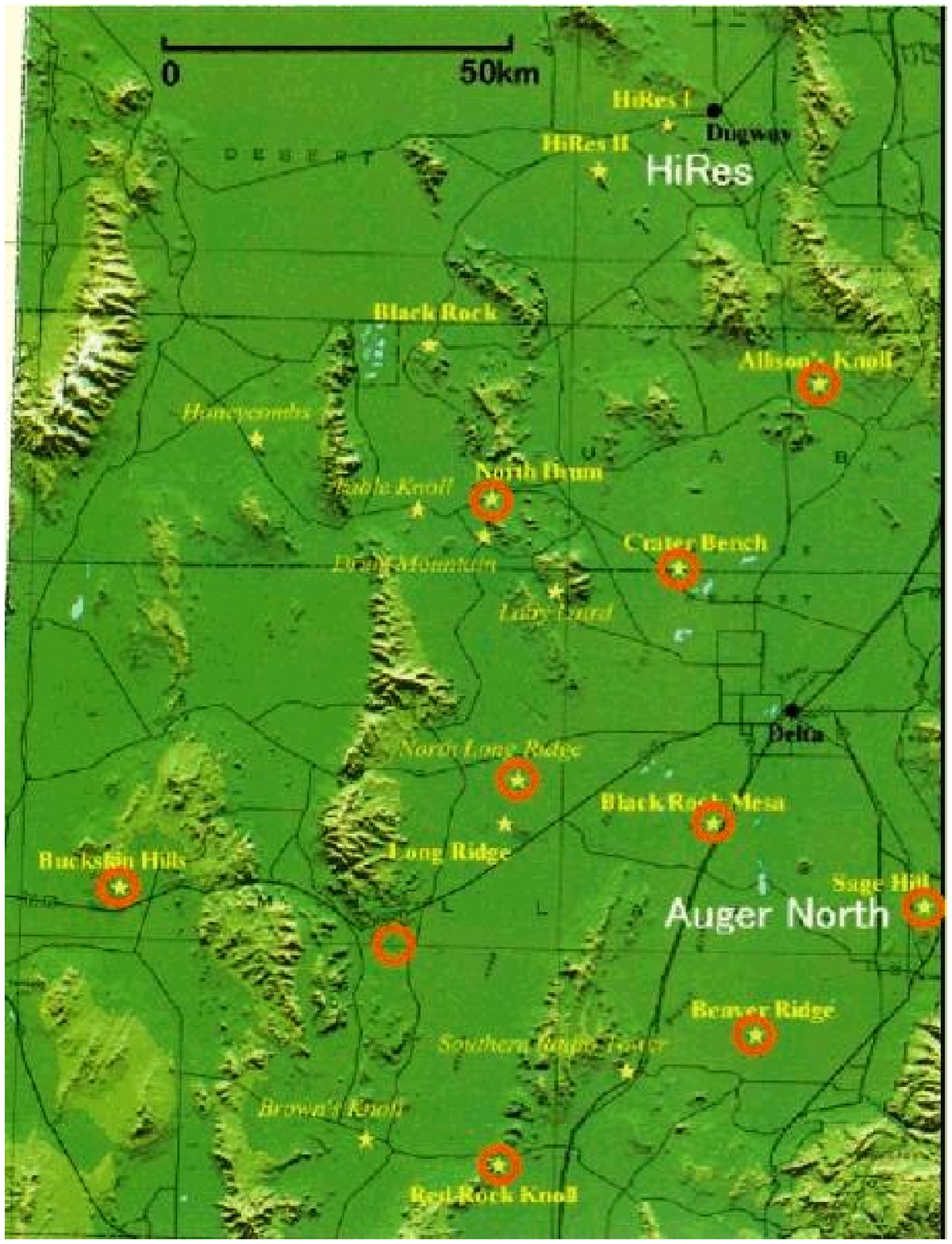}
\caption{\label{fig-sites}TA site arrangement.}
\end{minipage}
\hspace{10mm}
\begin{minipage}{0.35\linewidth}
\includegraphics[width=\linewidth,bb=0 150 567 650]{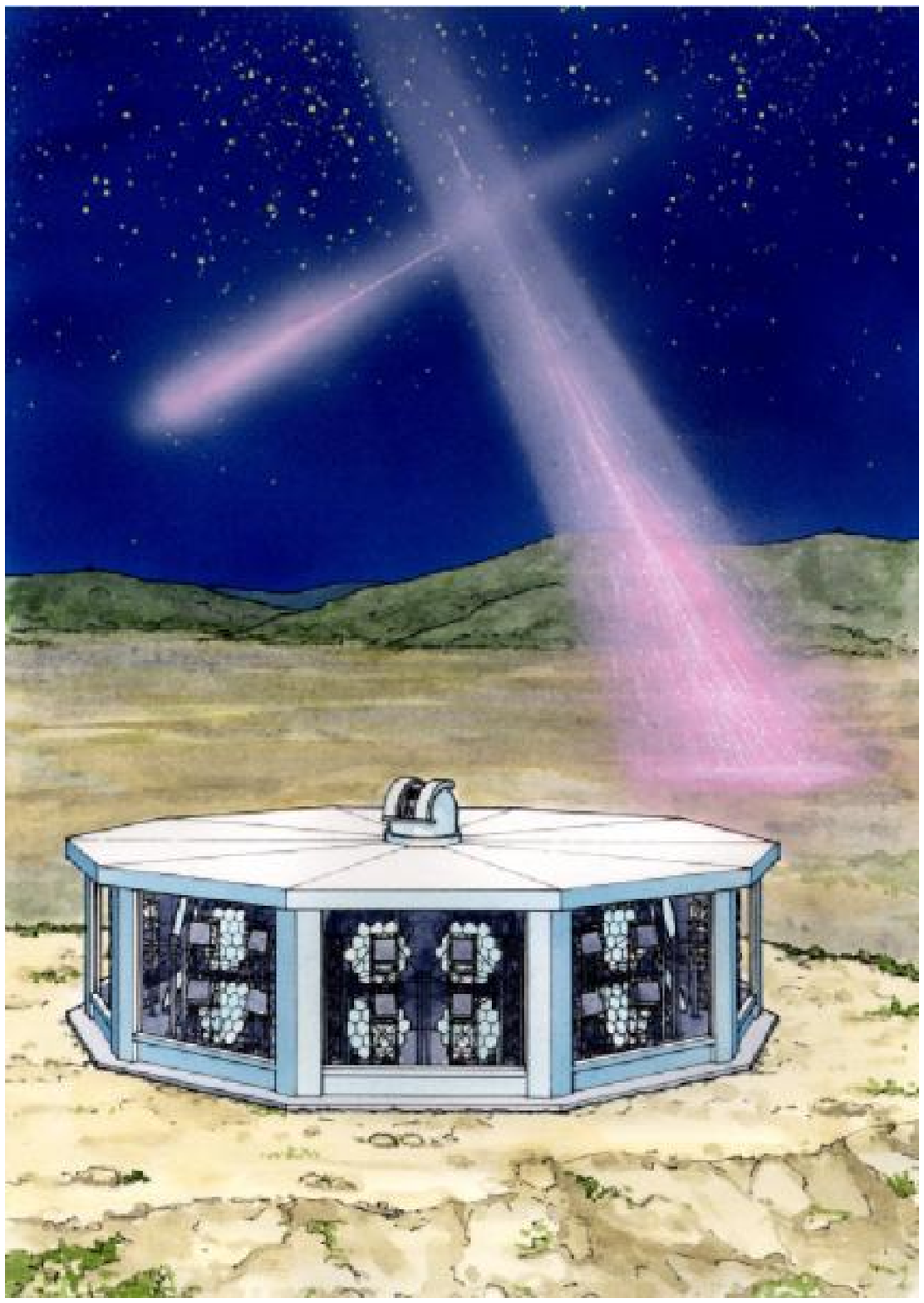}
\caption{\label{fig-stat_t01}TA detector station.}
\end{minipage}
\end{center}
\end{figure*}

The TA detector (Figs.\ref{fig-sites} and \ref{fig-stat_t01}) 
has been designed in order to
investigate
the origin of the highest energy cosmic rays
~\cite{sasaki97,TA00}.
For this purpose, the detector is required to obtain statistics 
of a much greater number of events
than the rate of one super-GZK event per year that the AGASA is able to detect. 
It should also
provide particle identification as well as accurate determination
of the direction of the primary cosmic ray to test the source models at a high confidence level.
The basic measurement principle use a huge effective aperture
to detect the image of fluorescence light 
produced by air showers and from this  
the longitudinal shower development can be reconstructed. 

The TA detector consists of 10 observational stations installed over an area at
about 30-40~km intervals as shown in Fig.~\ref{fig-sites}. 
Each station consists of 40 telescopes
with a 3~m-diameter f/1 mirror system on 2 layers of supports. 
256 2-inch photo multiplier tubes (PMTs) mounted
on the focal plane serve as pixels of the fluorescence sensor of each telescope.
Each PMT covers a visual field of angular aperture
1.1$^{\circ}\times$1.0$^{\circ}$.
We expect the
detection rate by the TA detector for events with energies beyond    the
GZK-cutoff to exceed that of AGASA by about 30 times. 
We note that this 
is similar to that expected for another proposed detector, 
the Pierre Auger Observatory (AUGER)~\cite{AUGER}.
The accuracy of
determining the energy and arrival direction for the highest energy events
are roughly 25\% and 0.6$^{\circ}$  respectively.

In order for the detector stations to have
good visibility, atmospheric transparency, no significant nearby sources
of light pollution, and be away from significant traffic, 
it is preferable
to install them 
on top of small mountains or hills, 
while still maintaining the arrangement in 30-40~km intervals. 
We have identified sites in the  south-west area of Delta
in southern Utah. 
The planned detector station array covers an area of  200~km$^2$.

\section{Neutrinos Deeply Penetrating the Atmosphere}
Neutrinos can produce air showers 
through interactions with the atmosphere.
The number of the shower depends on the particular interaction.
We consider here both
inelastic charged and neutral current interactions, which always produce
hadronic showers. 
Note however that
the case of charged current electron neutrino interactions,
in addition to the emerging electron, 
a pure electromagnetic shower
carrying a large fraction of the incoming particle energy is produced.

In our discussions of neutrino induced showers
we make use of
new calculations of the cross sections of charged-current and neutral
current interactions of neutrinos with nuclei
~\cite{GQRS:98,GQRS:96},
according to the CTEQ4-DIS (deep inelastic scattering) 
parton distributions~\cite{CTEQ:97}.
The CTEQ4-DIS parton distributions take account of new information about 
parton distributions within the nucleon~\cite{Quigg:97} using accurate
and extensive DIS data from the New Muon Collaboration (NMC)~\cite{NMC:95} and
DESY ep collider HERA~\cite{H1:95:96,ZEUS:95}, as well as new data from
E665~\cite{E665:96}. 

For a neutrino flux \( dI_{\nu }/dE_{\nu } \) interacting through a particular
process with differential cross section \( d\sigma /dy \), 
where \emph{y} is the fraction
of the incident particle energy transferred to the target, the event rate for
deeply penetrating showers can be obtained by a simple convolution:
\begin{eqnarray}
\label{eq:Rate}
Rate[E_{\nu}>E_{th}] &=& \nonumber \\
N_{A}\rho _{air}\int _{E_{th}}^{\infty } & dE_{\nu} & \int _{0}^{1} dy 
\frac{dI_{\nu }}{dE_{\nu }}(E_{\nu })\frac{d\sigma }{dy}(E_{\nu },y)
\epsilon (E_{sh})
,
\nonumber
\end{eqnarray}
 where \( N_{A} \) is Avogadro's number and \( \rho _{air} \) is the air density.
The energy integral corresponds to 
the primary neutrino energy \( E_{\nu } \) 
which is related to 
the shower energy \( E_{sh} \),
the relationship being different for each interaction.
\( \epsilon  \) is the detector acceptance,
a function of shower energy, which corresponds to the volume and solid angle
integrals for different shower positions and orientations with respect to the
detector. 
The function is different for both showers induced by charged current electron
neutrino interactions and those arising in neutral current or muon neutrino
interactions. 
This is because hadronic and electromagnetic showers have different
particle distribution functions. 
For \( (\nu _{e}+\overline{\nu }_{e})N \)
charged current interactions, we take the shower energy to be the sum of hadronic
and electromagnetic energies, \( E_{sh}=E_{\nu } \). For \( (\nu _{\mu }+\overline{\nu }_{\mu })N \)
charged current interactions and for neutral current interactions, we take the
shower energy to be the hadronic energy, \( E_{sh}=yE_{\nu } \). 
We take the inelasity $y$ to be a function of $E_{\nu}$
as in reference~\cite{GQRS:96},
while \( <y>\sim  \)0.25 does not depend strongly on the primary
energy beyond \( 10^{16} \)~eV in both 
charged and neutral current interactions.

\begin{figure}[htb]
\begin{center}
\includegraphics[width=0.9\linewidth]{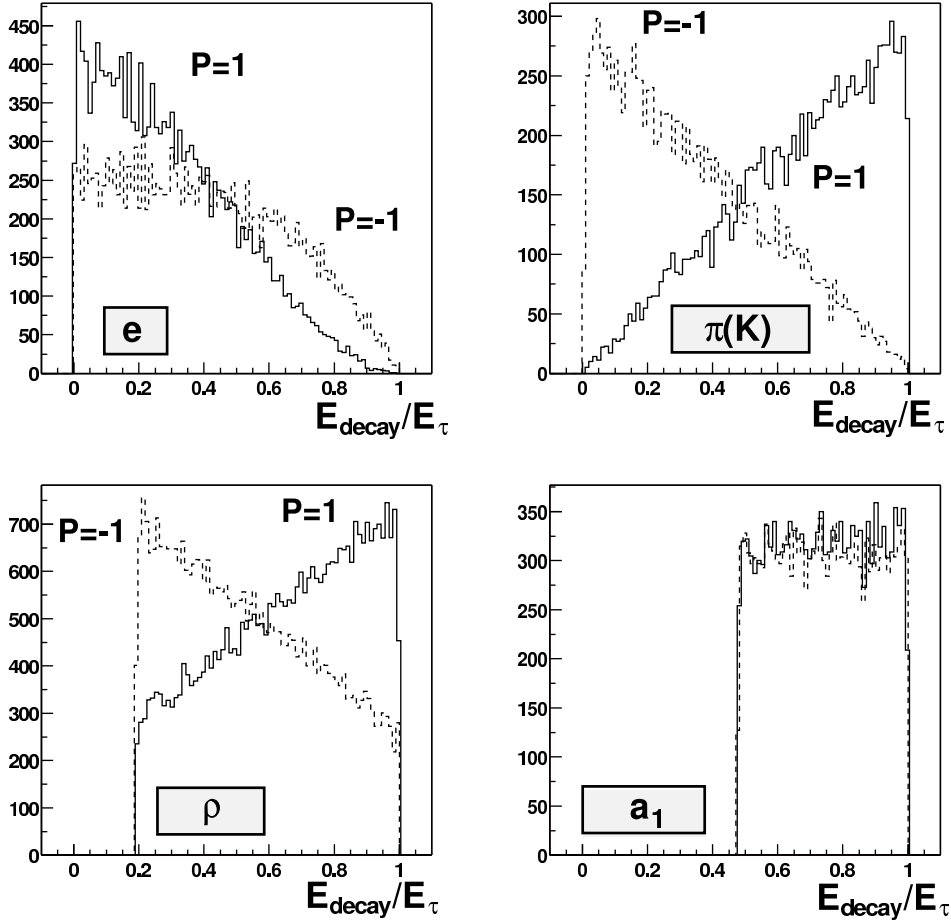}
\caption{
Distributions in visible decay energy divided by E$_{\tau}$
corresponding to values of tau polarization +/- 1
for
$\tau\rightarrow e\nu\bar{\nu}$,
$\tau\rightarrow\pi(K)\nu$,
$\tau\rightarrow\rho\nu$,  and
$\tau\rightarrow a_{1}\nu$.
\label{fig:taudecay-x}}
\end{center}
\end{figure}

In the case of $\nu_{\tau}$, we must also take into account the decay length of secondary tau leptons,
subsequent tau lepton decay, and a further energy input into 
the air shower from decaying electrons, photons, and hadrons.
The tau decay branching fractions and tau polarization effect 
are reasonably well accounted for simply by assuming that the energies of neutral mesons  
fully contribute to the electromagnetic showers.
The adopted decay branching fractions and the momentum spectra of the decay with
a polarization of $\pm$1 
in the colinear approximation in the laboratory frame are shown in 
Table~\ref{tab:taudecay} and Fig.~\ref{fig:taudecay-x} respectively. 
The $\tau$ decay length is often long enough that two subsequent energy inputs 
into the air showers can be seen in the field of view.

\begin{table}[htb]
\begin{center}
\caption{
Tau decay branching fractions from Particle Data Group (PDG)
~\cite{bib:PDG2000}
and used in the simulation for events in TA.
The branching fractions in the simulation should include
$\pi^{0}$ contributions.
\label{tab:taudecay}
}
\vspace{0.2cm}
\begin{tabular}{|c|c|c|}
\hline
Tau decay modes   &  PDG B.F. (\%)  &  B.F. (\%) in the simulation \\
\hline
$\tau\rightarrow \mu\nu\bar{\nu}$ & 17.37$\pm$0.07 & 17\\
$\tau\rightarrow   e\nu\bar{\nu}$ & 17.83$\pm$0.06 & 18\\
$\tau\rightarrow\pi(K)\nu$        & 11.79$\pm$0.12 & 13\\ 
$\tau\rightarrow\rho\nu$          & 25.40$\pm$0.14 & 37 ($\ge$ 1~neutral) \\
$\tau\rightarrow a_{1}\nu$        &  9.49$\pm$0.11 & 15 (all 3 prongs) \\
$\tau\rightarrow h h h \ge\pi^{0}\nu$ &  4.49$\pm$0.08 &              \\
\hline
\end{tabular}
\end{center}
\end{table}

Simulations for a given primary cosmic ray (electron neutrino, muon neutrino,
tau neutrino, and proton) were performed at fixed energies. 
The shower energy for neutrinos depends
on the generation, as described above. 
For each shower energy, the mean depth of the proton
shower maximum was determined from simulations~\cite{Gaisser:93}. For each primary
particle at each energy, the mean of the interaction length \( X_{1} \) was
determined from the above interaction cross sections of neutrinos with nuclei
or of protons,  \( X_{1} \)=83.1\( (E/\mbox{GeV})^{-0.052} \) g/cm\( ^{2} \)
~\cite{Honda:93}.
The shower energy determines the shower size at a maximum, 
\( N_{max} \)\cite{Baltrusaitis:85}.
Given \( N_{max} \), a slant depth at the maximum shower size
\( X_{max} \), and \( X_{1} \), the complete longitudinal
profile can be described by the Gaisser-Hillas function~\cite{Gaisser-Hillas:77}.
The Nishimura-Kamata-Greisen (NKG) lateral distribution function
~\cite{Kamata-Nishimura:58,Greisen:56}
normalized with the Gaisser parameterization has been used to find the total number
of electrons and positrons in hadronic (electromagnetic) cascade showers in order to
determine the location where fluorescence is produced. 
In our simulations we took into
account the fluctuations of the first interaction depth, impact point, 
and directional angles of air shower cores,
with appropriate distributions, but not air shower
size fluctuations. 

An important factor that must be calculated is
the light yield arriving at the detector site. 
For this the Rayleigh and Mie
scattering processes were simulated, with full account taken of the spectral
characteristics of the light. 
The isotropically emitted fluorescence light,
as well as direct and scattered Cerenkov light, is propagated, 
and the night sky background noise was added to the signal. 
All processes that affect the overall optical efficiency,
mirror area and reflectivity, optical filter transmission, 
and PMT quantum efficiency factors were included in the light spectrum 
to give the photoelectron yield in
each PMT, due to both signal and noise.

\begin{figure}[hbt]
\begin{center}
\includegraphics[width=0.48\linewidth]{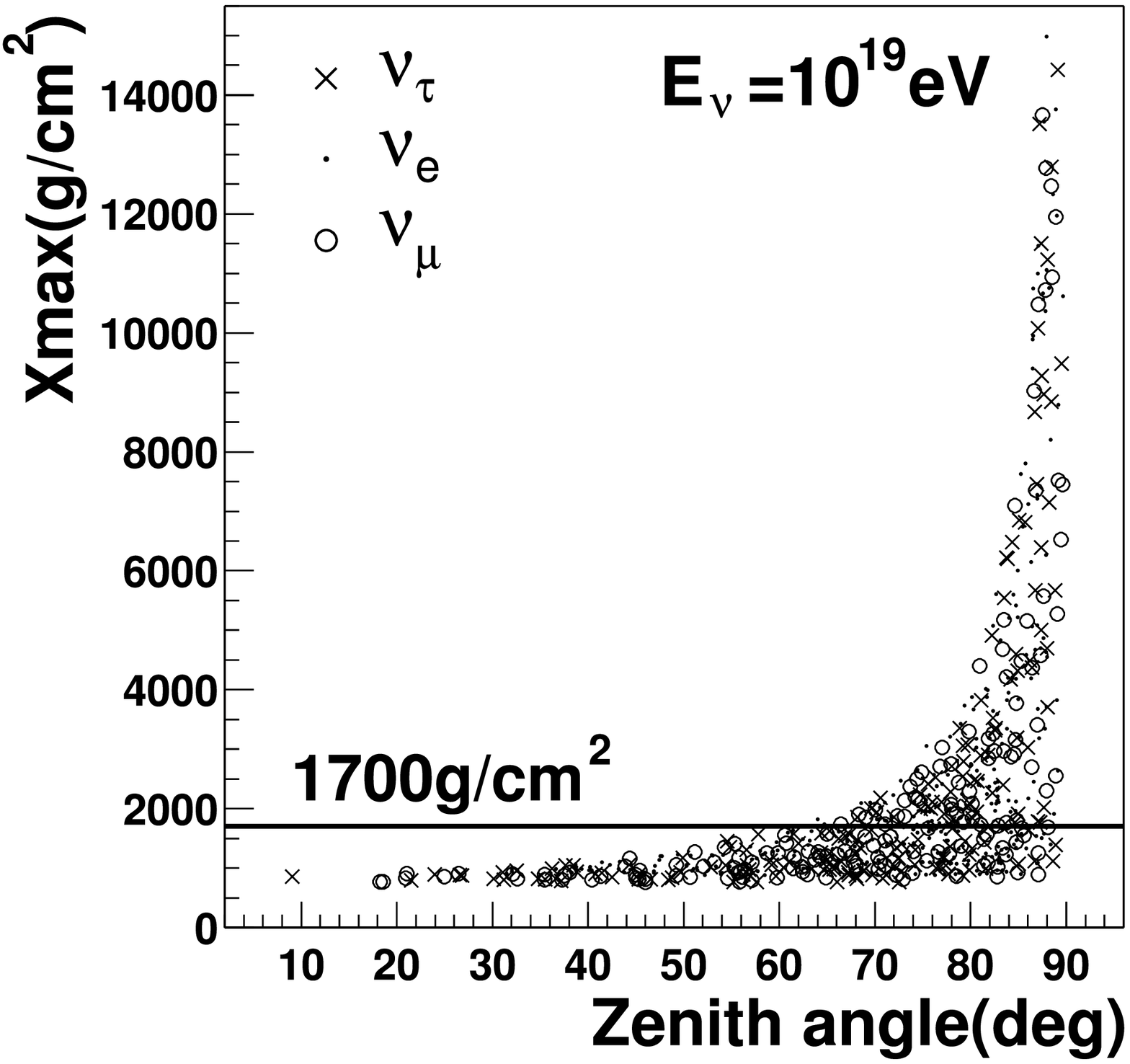}
\includegraphics[width=0.48\linewidth]{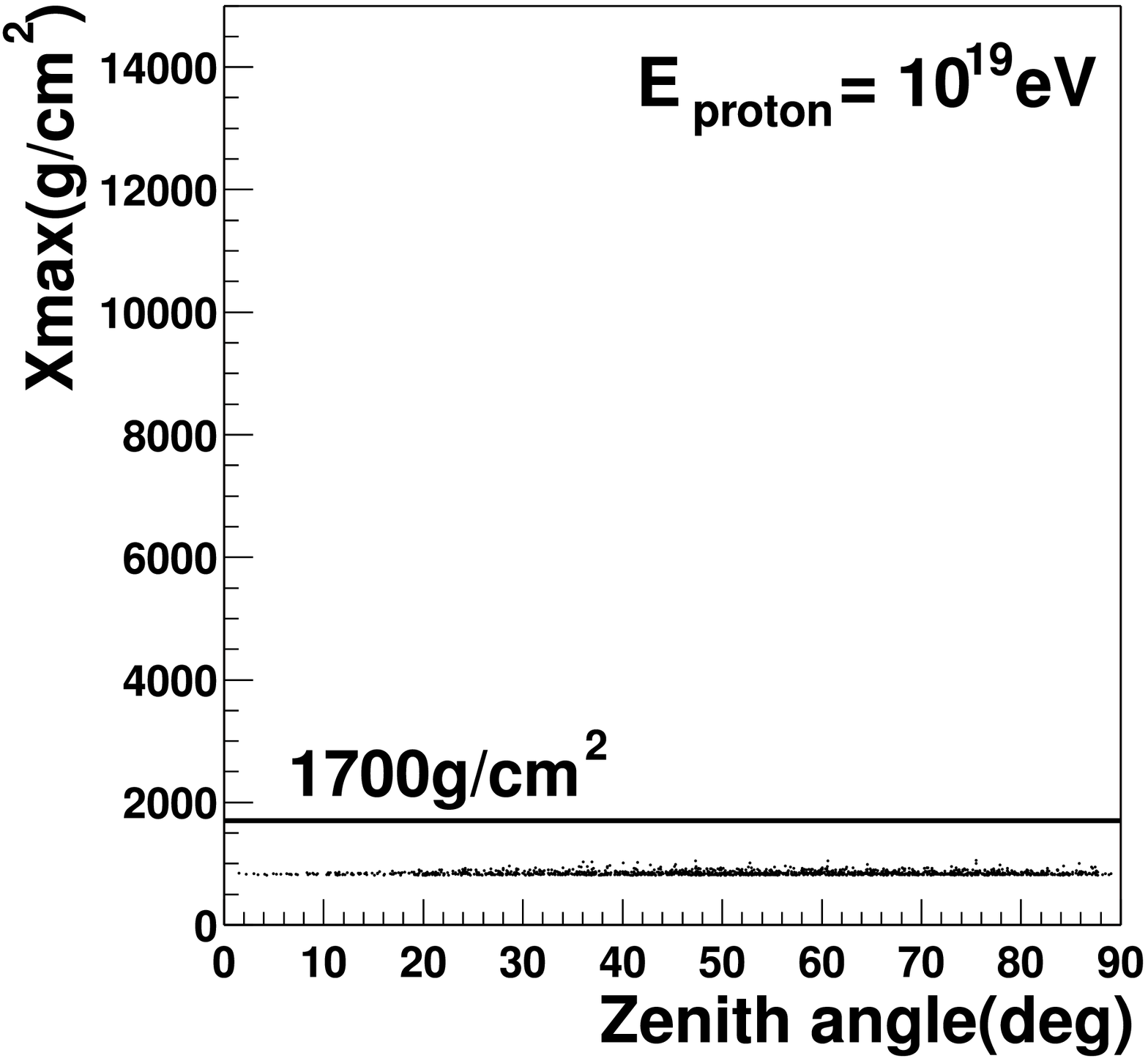}
\caption{
Shower slant depth distributions
for neutrino ({\it left}) and proton ({\it right})
induced events with energies of 10$^{19}$~eV 
from Monte Carlo simulations
showing the cut of 1700~g/cm$^2$, 
by which neutrino events can be distinguished from proton events.
\label{fig:xmax-zenith}}
\end{center}
\end{figure}

\begin{figure}[htb]
\begin{center}
\includegraphics[width=0.8\linewidth,bb=0 0 567 537]{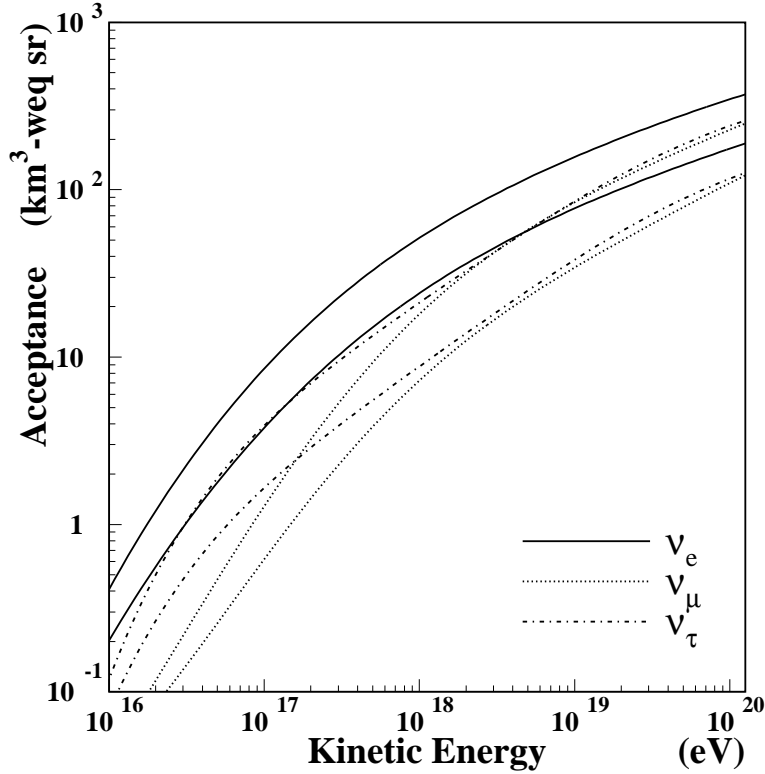}
\caption{\label{fig-agn-neut-accep}
Acceptance of the TA detector (1 stations) to a neutrino induced
air shower. Volume units are km\protect\( ^{3}\protect \) of water equivalent.
The higher and lower curves correspond to events after preselection
and proton rejection cuts respectively.}
\end{center}
\end{figure}

A ``fired'' PMT is defined such that its instantaneous photoelectron
current is greater than the 4\( \sigma  \) noise level of the night sky background.
We preselected events such that
at least one of the 10 stations contains at least 6 firing PMTs. 
To ensure track quality, we cut events for which shower maxima were not
viewed by any station. 
Finally, to reject proton background events, we selected only
events with \( X_{max}>1700 \) g/cm\( ^{2} \). 
Once the selection cuts were made the detector aperture
could be determined and the acceptances calculated by
multiplying the appropriate interaction length of a neutrino with a nucleon. 
Figure~\ref{fig:xmax-zenith} shows the shower slant depth distributions 
from Monte Carlo simulations
by which neutrino events can be distinguished from proton events.
The results are shown in Fig.\,\ref{fig-agn-neut-accep} for electron, muon, and tau
neutrinos as a function of the primary energy.
The acceptances of the TA for high-energy neutrinos deeply penetrating air 
compare well with those of IceCube and AUGER
~\cite{bib:sasaki00,GQRS:96,bib:Capelle98}
in lower and higher energy regions respectively,
even if we take into account the 10~\% duty factor of TA,
which is a rate of the observation time at moonless night in good weather.

\section{Earth-skimming Tau Neutrinos}
\begin{figure}[htb]
\begin{center}
\includegraphics[width=0.8\linewidth]{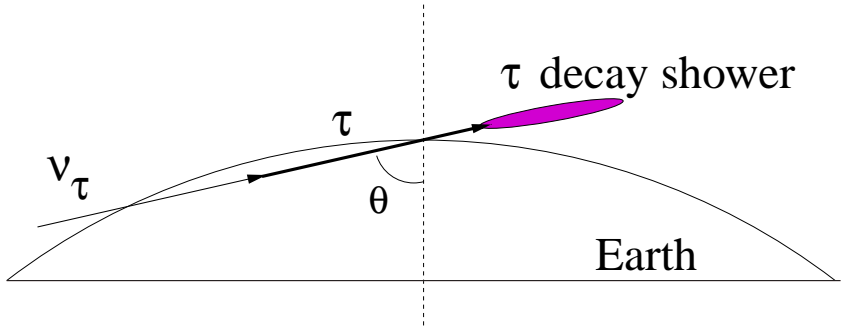}
\caption{
A schematic picture of Earth-skimming tau neutrino events.
\label{fig:skm-schematic}}
\end{center}
\end{figure}

Very high energy neutrinos penetrate the Earth and convert to charged leptons 
which then travel through the Earth.
This sequence is illustrated for an event with a nadir angle $\theta$
in Fig.~\ref{fig:skm-schematic}.
We define the critical angle $\theta_c$ such that
the chord thickness at the nadir angle $\theta_c$ corresponds to 
the charged current interaction length
$L^{\nu}_{CC}(E_{\nu})$ 
determined by the interaction cross section for a 
neutrino traveling with energy $E_{\nu}$.
For nadir angles smaller than $\theta_c$, neutrinos are shadowed by the 
Earth, and for larger nadir angles,
they rarely interact to produce charged leptons.
Table~\ref{tab:intlen} shows
charged current cross sections, interaction lengths, and 
$90^{\circ}-\theta_c$ at various neutrino energies.
For neutrino energies above 10$^{17}$~eV, 
$90^{\circ}-\theta_{c}$ is small and both
the neutrinos and the created leptons travel essentially horizontally.

\begin{table}[htb]
\begin{center}
\caption{
Charged current (CC) cross sections, interaction lengths 
and  90$^{\circ}-\theta_{c}$
(see text)
for $\nu$N interactions
for the CTEQ-DIS distributions~\cite{GQRS:96}.
\label{tab:intlen}
}
\vspace{0.2cm}
\begin{tabular}{|c|c|c|c|}
\hline
$E_{\nu}$ (eV)  &  
$\sigma^{\nu}_{CC}$ ( $10^{-33}$ cm$^2$) 
&  $L^{\nu}_{CC}$ ($10^7$g/cm$^2$) 
& $90^{\circ}-\theta_{c}$ ($^{\circ}$) \\
\hline
10$^{16}$ & 1.8 &  94 & 16   \\
10$^{17}$ & 4.8 &  35 & 5.9  \\
10$^{18}$ & 12  &  14 & 2.3  \\
10$^{19}$ & 30  &  5.5 & 0.89 \\
10$^{20}$ & 71  &  2.4 & 0.35 \\
\hline
\end{tabular}
\end{center}
\end{table}

\begin{figure}[htb]
\begin{center}
\begin{minipage}[t]{0.48\linewidth}
\includegraphics[width=\linewidth]{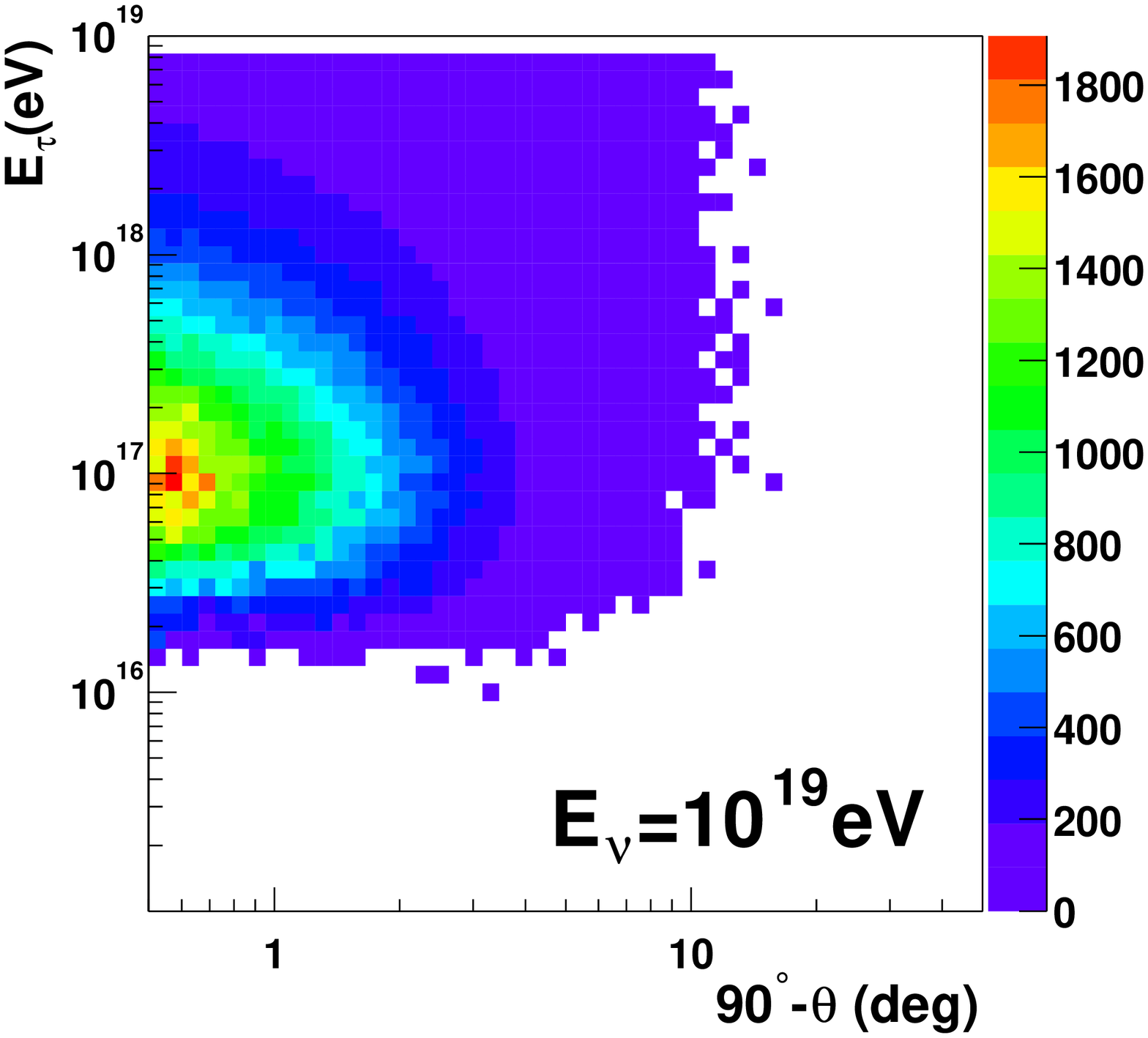}
\end{minipage}
\hspace{1mm}
\begin{minipage}[t]{0.48\linewidth}
\includegraphics[width=\linewidth]{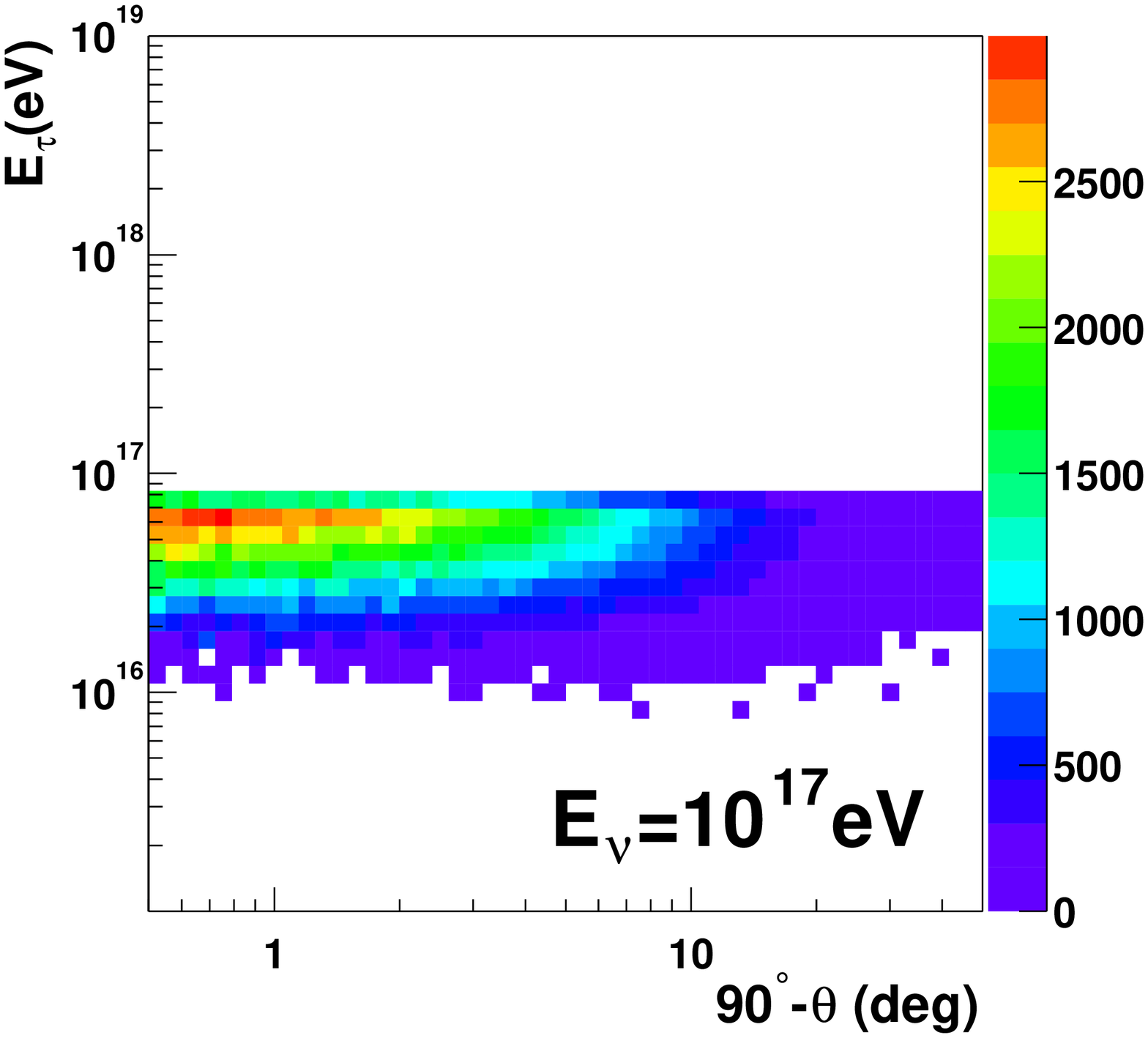}
\end{minipage}
\caption{
Distribution on the plane of energy E$_{\tau}$ and  
90$^{\circ}-\theta$ 
of tau leptons 
that exit the Earth in the case of a primary neutrino energy of 
10$^{19}$~eV ({\it left}) and 10$^{17}$~eV ({\it right}).
\label{fig:kern}}
\end{center}
\end{figure}

Once created, a charged lepton loses energy through Bremsstrahlung, 
pair production, and photonuclear interactions.
Electrons lose their energy too quickly in the Earth 
and hence cannot be detected by fluorescence.
Assuming that a lepton loses energy uniformly and continuously,
its energy loss can be parametrized by:
$
dE_{l} / dz = - \beta_{l} \rho E_{l} ~~,
$
where 
the Earth's density $\rho$ can be simplified to be 2.65~g/cm$^3$
uniformly for small nadir angles.
The values of the constant $\beta_{l}$ are
$
\beta_{\mu}  \sim 6.0 \times 10^{-6} cm^{2}/g 
$ and
$
\beta_{\tau} \sim 0.8 \times 10^{-6} cm^{2}/g
$
for muons and tau leptons respectively
for the energies of interest here~\cite{Lipari-Stanev:91}.
Hence, at the Earth's surface muons and tau leptons can
travel 1.5~km and 11~km, respectively, before losing a decade in energy.
Therefore, tau neutrinos contribute dominantly to
Earth-skimming events observed above the Earth's surface.
Figure~\ref{fig:kern} shows
distributions on the plane of energy $E_{\tau}$ and 
$90^{\circ}-\theta$ of tau leptons that exit the Earth 
from primary neutrinos with energies of 
10$^{19}$~eV and 10$^{17}$~eV.

The survival probability $P_{svv}$ for a tau losing energy as it moves through the
Earth is described by the energy loss $dE_{\tau}/dz$ as above,       
and 
$$
\frac{dP_{svv}}{dz} = - \frac{ P_{svv}}{c\tau_{\tau} E_{\tau}/m_{\tau}} ~~,
$$
where $c$ is the speed of light and $m_{\tau}$ and $\tau_{\tau}$ are tau
lepton's rest mass and lifetime respectively.
These can be solved and the survival probability given by
$$
P_{svv} = \exp{[\frac{m_{\tau}}{c\tau_{\tau}\beta_{\tau}\rho}
                     (\frac{1}{E_{\nu}}-\frac{1}{E_{\tau}})]} ~~.
$$
For tau leptons, this factor plays a significant role.
The competition between $L^{\nu}_{CC}$ and $P_{svv}$
makes the peak of the distribution at the 
survived tau lepton energy of $10^{17}$~eV
in Fig.~\ref{fig:kern}, which is independent of the
primary tau neutrino energy.
Due to this effect, 
lower energy sensitivity of an air fluorescence telescope
makes a key role to detect the Earth skimming tau neutrinos.

The flux of survived taus with energy $E_{min} < E_{\tau} < E_{max}$ is
$$
\Phi_{\tau} = \frac{ln(E_{max}/E_{min})}{2 R \beta_{\tau}\rho} \Phi_{\nu}~~,
$$
where $\Phi_{\nu}$ is the flux of neutrinos.
As a result, the flux in any given decade in tau energy is
$\Phi_{\tau} = 8.5\times 10^{-4} \Phi_{\nu}$,
that is, 1 in every 1200 neutrinos that skims the Earth  
emerges as a tau lepton with the required energy~\cite{bib:Feng01}.
This remarkably simple and robust statement is quite useful
because it is independent of the neutrino energy and microscopic details.
We have examined our Monte Carlo simulation results by comparison with 
this analytic estimate.

\begin{figure}[htb]
\begin{center}
\includegraphics[width=0.8\linewidth,bb=0 0 567 537]{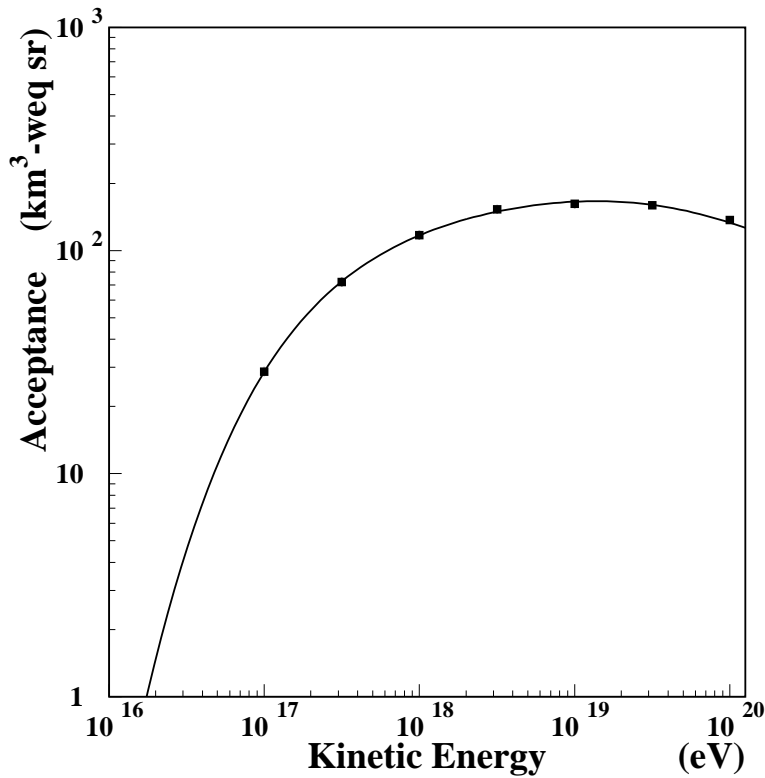}
\caption{
Detection acceptance estimate (in km$^3$-water-equivalent steradian)
for Earth-skimming tau leptons through
their decays to electromagnetic and/or hadronic air showers with 
only one TA detector station.
\label{fig:skmtau-acceptance}}
\end{center}
\end{figure}

The tau decay length is given by 
$                
L_{\tau} = c \tau_{\tau} (E_{\tau}/m_{\tau}) \sim$
 4.9~km ($E_{\tau}/10^{17}$~eV)
, using 
the lifetime $c \tau_{\tau}= 87.11~\mu$m 
the tau mass $m_{\tau}= 1777.03$~MeV 
from PDG~\cite{bib:PDG2000}.
The complete treatment for tau decays can be performed 
as described in the previous section.
Effective detection apertures and acceptances
have been estimated for Earth-skimming tau events
changing the neutrino energies 
from the detailed Monte Carlo simulation.
Figure~\ref{fig:skmtau-acceptance}
shows the effective detection acceptances
using only one TA station.
The tau decay length as a function of the tau lepton energy
which is observed within the view of the telescope
makes the detection acceptance slightly decrease 
as increasing the kinetic energy of Earth skimming tau neutrino above 
the primary energy of 10$^{19}$~eV.

\section{Detection Event Rates}
Models that assumes shock
acceleration in the Active Galactic Neuclei (AGN)
cores predict relatively flat fluxes up to energies
of about 10\( ^{15} \)~eV~\cite{bib:SS95}. 
The GeV to TeV gamma ray emissions observed in AGN corresponds to the
blazar class. 
Most recent models for the proton blazars site the acceleration
in the jets themselves. 
We use the prediction of reference~\cite{bib:Mannheim95}, 
which illustrates that the emitted neutrinos may extend well into the EeV
region.
Annual event rates have been calculated
for the TA detector for neutrino
induced air showers with fluxes 
from GRB model~\cite{bib:Waxman95} and  
Greisen neutrino~\cite{bib:Engel01}  models
as well as from AGN-core~\cite{bib:SS95} and
AGN-jet~\cite{bib:Mannheim95},
assuming a duty cycle of 10\%, 
and are shown in Table~\ref{table-agn-neut-rate}.
The testability of the bottom-up scenario models is statistically simple  
even when considering only shower events deeply penetrating the air.

We also compare the detectability of Earth-skimming tau neutrino events
with the downward events described above.
Table~\ref{table-agn-neut-rate} also
lists annual event rates for neutrino induced air-showers 
from various models including top-down scenario sources
of shower events deeply penetrating the air and Earth-skimming events.
The duty factor of 10\% is assumed as usual.
The enhancement of the statistics by considering
Earth-skimming events is significant,
particularly in the ultra-high energy region.
Hence it is feasible to
test the top-down scenario models such as topological defects and Z-bursts
in a statistically significant way by considering both downward 
and upward neutrino-induced air-shower events
in complementary energy regions.

\begin{table}[t!]
\begin{center}
\caption{\label{table-agn-neut-rate}
Annual event rates in the TA detector for neutrino
induced air showers with fluxes 
from AGN-core~\cite{bib:SS95}, AGN-jet~\cite{bib:Mannheim95},
GRB model~\cite{bib:Waxman95},  
Greisen neutrino~\cite{bib:Engel01},
topological defects~\cite{bib:Sigl97}, and
Z-bursts~\cite{bib:Yoshida98} models
(see text).
A duty factor of 10\% was assumed.}
\vspace{0.2cm}
\begin{tabular}{|c|ccccc|c|} 
\hline

  & CC$\nu_e$ & CC$\nu_\mu$ & CC$\nu_\tau$ & NC & ES$\nu_\tau$ & Total\\

\hline

 AGN-core &    8.9 &    1.4 &   4.2 &   1.7 &   3.1 &   19.3 \\
%
%
 AGN-jet  &    6.8 &    2.4 &   3.4 &   3.1 &  42.8 &   58.4 \\
 GRB      &   0.48 &   0.17 &  0.25 &  0.22 &  2.48 &   3.60 \\
 Greisen  &   0.52 &   0.25 &  0.30 &  0.32 &  3.21 &   4.59 \\
 TD       &   0.09 &   0.06 &  0.06 &  0.07 &  0.34 &   0.63 \\
 Z-Bursts &   0.46 &   0.33 &  0.34 &  0.40 &  1.19 &   2.72 \\



\hline
\end{tabular}
\end{center}
\end{table}

\section{Conclusions}
The next generation of cosmic ray detectors such as 
AUGER and the TA have target volumes of atmosphere
competitive with or better than that of IceCube and their
annual detection rates
assuming AGN-jet proton acceleration models are statistically sizable
~\cite{bib:sasaki00,bib:Capelle98}.
The TA using an air fluorescence technique is 
advanced as a method for discriminating primary neutrinos 
proton induced air showers.
Recently a novel strategy for detecting
Earth-skimming extremely high energy neutrinos
has been quantitatively proposed for which significant
improvements in detection can be made using the
TA detector compared to down-going neutrino detection~\cite{bib:Feng01}. 
Encouragingly, extremely high energy cosmic neutrino sources,
to say nothing of AGN proton acceleration models, such as
Greisen photo production~\cite{bib:Engel01},
topological defects~\cite{bib:Sigl97}, long-lived super heavy particles,
and Z-bursts~\cite{bib:Yoshida98}, 
may be experimentally tested with the TA.

\section{Acknowledgments}
We are indebted to my colleagues in the
TA Collaboration, especially those who
have contributed to the TA Design Report,
for assistance in the preparation of this paper.
We would like to especially acknowledge the contributions of
Prof.~T.~Matsuda and Prof.~C.~Fukunaga
for useful discussions.


\end{document}